\documentclass{elsart4}

\usepackage{amssymb}

\usepackage{graphicx}
\usepackage{amssymb}
\journal{Physica B}

\begin{document}

\begin{frontmatter}

\date{7 August 2011}



\title{Novel method for photovoltaic energy conversion using surface acoustic waves in piezoelectric semiconductors}

\author{Victor M.~Yakovenko}  
\address{Department of Physics, University of Maryland, College Park, MD 20742-4111, USA}

\vspace{-0.7cm}
\begin{abstract}
This paper presents a novel principle for photovoltaic (PV) energy conversion using surface acoustic waves (SAWs) in piezoelectric semiconductors.  A SAW produces a periodically modulated electric potential, which spatially segregates photoexcited electrons and holes to the maxima and minima of the SAW potential.  The moving SAW collectively transports the carriers with the speed of sound to the electrodes made of different materials, which extract electrons and holes separately and generate dc output.  The proposed active design is expected to have higher efficiency than passive designs of the existing PV devices and to produce enough energy to sustain the SAW. 
\vspace{-0.5cm}
\end{abstract}

\begin{keyword} 
surface acoustic waves
\sep
piezoelectric semiconductors
\sep
photovoltaics
\sep 
solar energy
\sep
photocurrent
\sep
quantum efficiency

\PACS
84.60.Jt (Photoelectric conversion)
\sep
88.40.jm (Thin film III-V and II-VI based solar cells)
\sep
85.50.-n (Dielectric, ferroelectric, and piezoelectric devices)

\end{keyword}

\end{frontmatter}


\paragraph*{Introduction.}

This paper presents a theoretical proposal \cite{Yakovenko-2009} for a new principle for photovoltaic (PV) energy conversion using surface acoustic waves (SAWs) in piezoelectric semiconductors.  As shown below, SAWs can spatially segregate photogenerated electrons and holes and transport them collectively with the speed of sound to the electrodes.  The principal new element of this proposal is active management of the photogenerated electrons and holes, as opposed to passive in the existing PV devices, which would increase efficiency of energy conversion.  The proposed new principle can be implemented in different ways, so this paper only presents the general concept.   It can be also combined with other well-known methods for improving efficiency, such as optical focusing of sunlight and multiple energy gaps for better spectral coverage.  

In general, PV energy conversion consists of 4 steps:
\begin{enumerate}

\item Absorption.  Photons are absorbed across an energy gap and generate electrons above and holes below the energy gap.

\item Segregation.  The photogenerated electrons and holes are spatially separated by an electric field to prevent their recombination.

\item Transportation.  Electrons and holes are transported to the collecting electrodes, which are usually located on the periphery of a device.

\item Extraction.  Electrons and holes are extracted to the electrodes of an output electric circuit.

\end{enumerate}
We briefly review the basics of SAWs and then describe how these steps can be implemented using SAWs.

\vspace{-3mm}
\paragraph*{SAW basics.} 

Acoustic modes represent wave-like propagation of crystal lattice distortions.  There are modes that propagate in the bulk, but there are also acoustic modes that propagate along the surface of a crystal.  The amplitude of the latter modes exponentially decays into the bulk, so the modes are localized near the surface and are called SAWs.  All modes of acoustic waves are characterized by distinct dispersion relations $\omega_s(k)$ between the wave number $k$ and the frequency $\omega_s$ of the wave.  Typically, SAWs are generated using an interdigitated transducer (IDT), which consists of interpenetrating metallic fingers placed on the surface of a material.  The spacing $\lambda$ between the fingers determines the wave number $k=2\pi/\lambda$ of the SAW.  When the frequency $\omega$ of an ac voltage applied to the interdigitated fingers matches the dispersion relation $\omega_s(k)$ of the SAW for the fixed $k=2\pi/\lambda$, the SAW is generated.  In Ref.\ \cite{PRL-1997}, the SAW was utilized with the following parameters: frequency $f=840$~MHz, wavelength $\lambda=3.4$~$\mu$m, and speed $v\approx3$~km/s.  SAWs are widely used commercially, e.g.,\ in cell phones,  as frequency filters and delay lines.

\begin{figure}
\centering
\includegraphics[width=0.35\linewidth]{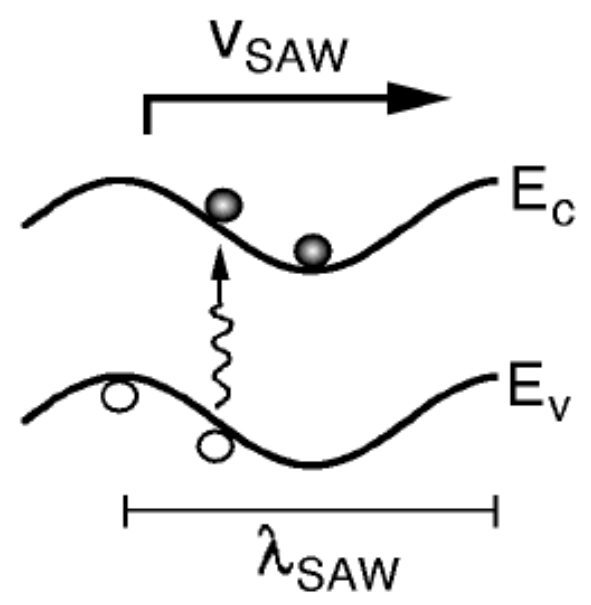}
\caption{Absorption of a photon across the energy creates an electron in the conduction band and a hole in the valence band, which then relax to the minima and maxima of the periodic electric potential produced by the SAW (see Ref.~\cite{PRL-1997}).}
\label{fig:SAW}
\end{figure}

In piezoelectric semiconductors, such as GaAs, the two elements, Ga and As, have opposite electric charges because of their different electron affinities, thereby forming a dipole moment between them.  A propagating SAW causes periodic deformation of the lattice, which results in a periodically modulated dipole moment due to displacement of the oppositely charged Ga and As atoms.  The result is a periodically modulated electric potential, which shifts the energy gap of the semiconductor up and down in energy, as shown in Fig.~\ref{fig:SAW}.  

\vspace{-3mm}
\paragraph*{Spatial segregation of photogenerated electrons and holes by the SAW (steps 1 and 2).} 

Let us consider a pure (undoped) piezoelectric semiconductor, where a SAW is generated by an IDT.  In step 1, photons are absorbed across the energy gap and create electrons above and holes below the energy gap.  Then, in step 2, the electrons are pulled to the minima of the periodically modulated electric potential of the SAW, and the holes to the maxima, as shown in Fig.~\ref{fig:SAW}.  Thus, the electrons and holes go to the troughs and crests of the SAW and spatially separate from each other.  The spatial separation is forced by the periodically modulated electric field exerted by the SAW on the electrons and holes, which can be as strong as 10~kV/cm depending on the SAW amplitude \cite{PRL-1997}.  Once the spatial separation of electrons and holes is achieved, their recombination is strongly suppressed, and the traveling wave of stored electric charge can persist for quite a long time.  This was experimentally demonstrated in Ref.\ \cite{PRL-1997}.  Thus, step 2 of the PV energy conversion is accomplished in the whole volume where the SAW is present.  Moreover, this task is accomplished in a pure crystal without p and n doping.  The absence of doping simplifies technological procedures and eliminates traps and scattering centers present in conventional p-n junctions.

\vspace{-3mm}
\paragraph*{Collective transport of photogenerated electrons and holes by the SAW (step 3).} 

Once electrons and holes are segregated and captured by the SAW, the next step 3 is to transport them to external electrodes.  This step is also accomplished by the SAW, which carries the electrons and holes like passengers on a high-speed train, as it propagates with the speed of sound along the surface of the crystal.  This collective transport is expected to be more efficient than diffusive one in passive PV devices.  It somewhat resembles the well-known Fr\"ohlich conductivity for charge-density waves, but here it constitutes the ambipolar transport, where both electrons and hole move in the same direction.  Collective transport of carriers by a SAW was theoretically envisioned in Ref.~\cite{Rakhmanov-1970} and experimentally demonstrated in Ref.~\cite{PRL-1997}.   In Ref.~\cite{PRL-1997}, the photogenerated electrons and holes were transported by the SAW from a laser spot to a point where the SAW potential was screened out by a metallic layer, so the electrons and holes recombined and emitted photoluminescence observed by a detector. Subsequent experiments \cite{APL-2004} demonstrated collective transport by the SAW at room temperature over a distance longer than 0.1 mm.  Ref.\ \cite{NatMat-2005} employed two SAWs propagating in perpendicular directions to create a checkerboard pattern of dynamical quantum dots for quantum computing.

\vspace{-3mm}
\paragraph*{Extraction of photocurrent from the collectively transported electrons and holes (step 4).} 

In the experiments \cite{PRL-1997,APL-2004,NatMat-2005}, the captured photon energy was eventually re-emitted, so they did not achieve PV energy conversion.  Extraction of photoexcited electrons and holes from a moving SAW into an external electric circuit is a challenging problem, because both carriers move in the same direction.  Thus, it is necessary to perform further separation of electrons and holes in order to extract them to negative and positive electric terminals.  In Refs.\ \cite{APL-2008} and \cite{JAP-2009}, this task was accomplished by lateral p and n doping on the opposite sides of the sample, transverse to the direction of SAW propagation.  The built-in electric field of this transverse p-n junction forced electrons and holes to move in the opposite directions along the wave fronts of the SAW toward the corresponding contacts.  The output photocurrent was measured in Refs.\ \cite{APL-2008} and \cite{JAP-2009}, and the external quantum efficiency $\eta$, which is the number of extracted  electron-hole pairs per incoming photon, was determined.  When the SAW amplitude exceeded a certain threshold, the photocurrent sharply increased, and the quantum efficiency as high as $\eta=85\%$ over the transportation distance of 0.3 mm was achieved \cite{JAP-2009}. 

\vspace{-3mm}
\paragraph*{The proposed PV device using SAWs.}

The results of Refs.\ \cite{APL-2008} and \cite{JAP-2009} are encouraging and demonstrate that a high quantum efficiency of photocurrent conversion can be achieved with electron-hole capture and transport by SAWs.  However, the goal of these papers was to make a sensitive photodetector for low photon fluxes, not a device for PV energy conversion.  For this reason, only the output electric current was measured, but not the output electric voltage.  Thus, the \textit{power} conversion efficiency, i.e., the ratio of the output electric power to the input photon power, is not known.  Moreover, the design employed in Refs.\ \cite{APL-2008} and \cite{JAP-2009} included batteries in the output circuit in order to improve photocurrent.  

The collecting p and n electrodes in Refs.\ \cite{APL-2008} and \cite{JAP-2009} were placed closely to form a narrow channel, so that the transverse distance traveled by the electrons and holes is short.  However, this design would not be practical for samples with an approximately square aspect ratio.  In this geometry, the electrons and holes would have to travel in the transverse direction with a drift velocity comparable to the SAW speed in order to be collected by the electrodes during the time of flight, which is not be realistic.  

\begin{figure}
\centering
\includegraphics[width=0.8\linewidth]{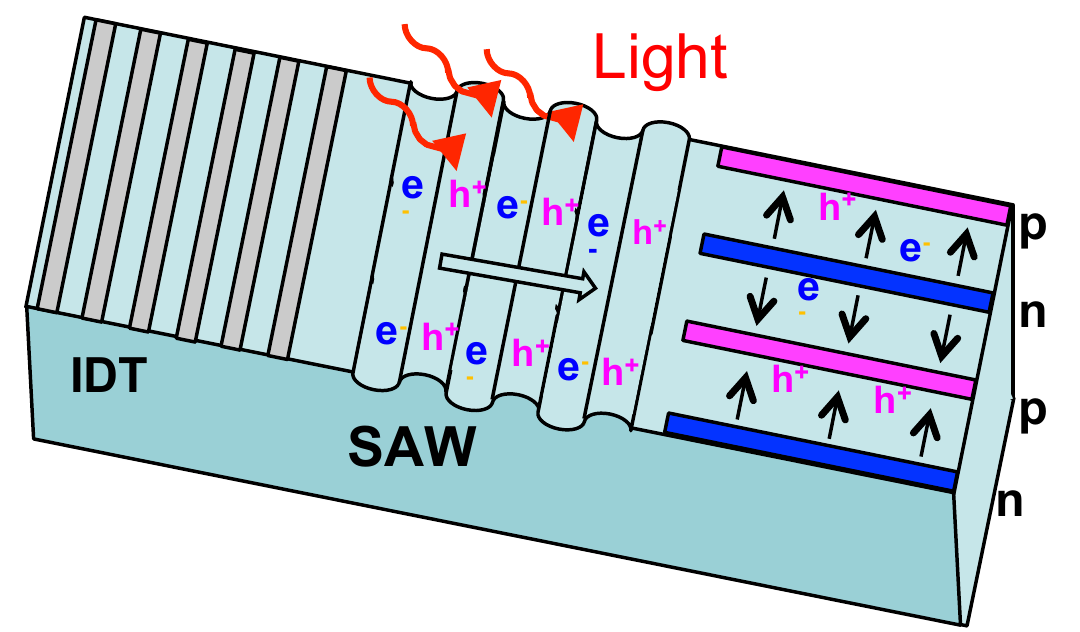}
\caption{Novel PV design using SAWs.  The IDT sends a SAW across a region exposed to light.  The photogenerated electrons and holes are captured to the minima and maxima of the SAW and get transported to the electrodes.  The electrodes are made from two different materials collecting p and n carriers.  The arrows show the transverse electric current of the electrons and holes unloading from the SAW.}
\label{fig:lateral}
\end{figure}

\begin{figure}
\centering
\includegraphics[width=0.8\linewidth]{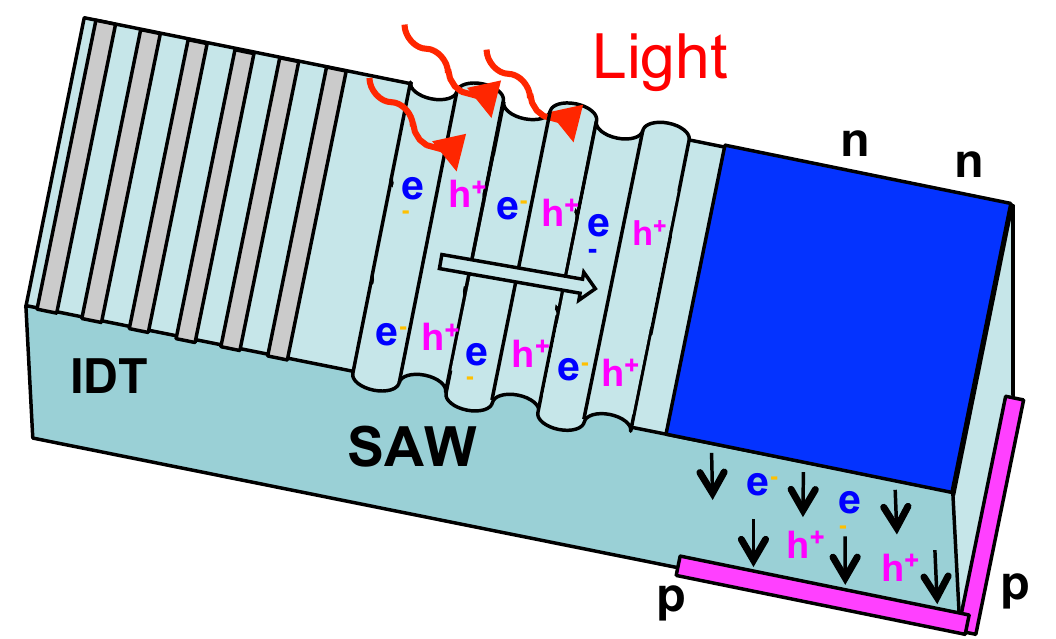}
\caption{The same as in Fig.~\ref{fig:lateral}, but with the electrodes placed above and below the sample.  The arrows show the vertical electric current of electrons and holes unloading from the SAW.}
\label{fig:vertical}
\end{figure}

To address this problem, alternative designs adapted for PV purposes are proposed in Figs.~\ref{fig:lateral} and \ref{fig:vertical}.  The p and n electrodes are placed either on top of the device in the alternating manner (Fig.~\ref{fig:lateral}) or vertically (Fig.~\ref{fig:vertical}).  In both cases, electrons and holes need to travel only short distances about few microns transversely or vertically.  The setup shown in Fig.\ \ref{fig:lateral} utilizes two sets of interdigitated leads extending in the direction parallel to the direction of SAW propagation.  The leads labeled as n and p are intended to collect electrons and holes, respectively, so that the dc current is generated in the output circuit.  The transverse arrows in Fig.\ \ref{fig:lateral} indicate the staggered built-in electric field between the p and n leads, which laterally separates electrons from holes.  The spacing between the leads should be short enough, and the leads should be long enough, so that the electrons and holes can reach the leads during the time of flight of the SAW along the leads.  Using the train analogy, the multiple leads act as platforms for parallel unloading of passengers (electrons and holes) from the train (the SAW) without stopping the train.  Another possibility is to orient the p and n leads in parallel to the SAW fronts with the spacing $\lambda_{\rm SAW}$ (not shown).

In the setup shown in Fig.~\ref{fig:vertical}, the p and n electrodes are placed above and below the sample, so the electrons and holes would travel vertically.  For a direct-gap semiconductor, such as GaAs, the absorption depth of light and the penetration depth of the SAW are both about few microns, so the vertical travel distance, indeed, can be short.  To prevent screening of the SAW periodic potential, the top and bottom electrodes can be patterned with multiple transverse slits to prevent redistribution of the screening electric charge.

The setup shown in Fig.~\ref{fig:vertical} resembles conventional PV design, where the p and n contacts are placed vertically over the whole surface area of the sample.  However, the unique feature of using SAWs for PV is separation of the surface areas where light is absorbed and where electrons and holes are extracted.  The light-collection area is directly exposed to light and does not have any obscuring top electrode, such as the transparent indium-tin oxide (ITO) or metallic wires.  This improves light collections efficiency and avoids usage of the relatively rare element Indium.  The electrodes are located on the side of the sample and occupy relatively small area compared with the light-collecting area.  The photoexcited electrons and holes are swept from the large light-collection area into the small electrode area by the SAW, which acts as an active concentrator.  The open-circuit voltage of a PV device is determined by the balance between photocurrent and dark current, which are proportional to the light-collection and electrode areas, respectively.  The open-circuit voltage for our setup is expected to be high because of the large ratio of these two areas.  Conventional PV devices, where the two areas are the same, do not have this additional degree of freedom.  Thus, the proposed SAW device is expected to have higher PV efficiency.

The p and n leads can be produced by doping or by metal-semiconductor contacts using two metals with different work functions (Fermi level positions) to obtain rectifying Schottky barriers of the opposite signs.  It is also possible to tune the properties of the Schottky contacts in piezoelectric semiconductors by applying local mechanical stress \cite{Wang-2010}.

\vspace{-3mm}
\paragraph*{Energy consumption for SAW generation.}

As a tradeoff of the active design, some energy has to be supplied to the IDT in order to sustain the SAW.  The energy of the required lattice deformation is expected to be relatively small compared with the PV energy extracted from the light.  The SAW amplitude and, thus, the rf energy consumption should be adjusted depending on the intensity of the incoming photon flux.  When the flux is high, the SAW amplitude should be increased to make the SAW potential deeper in order to accommodate more photogenerated electrons and holes, and vice versa.  Adaptive management of the rf power should reduce energy consumption.  Moreover, the overall efficiency depends on the scale of the device.  A single rf generator feeding many PV cells reduces energy consumption per unit area. 

It is possible, in principle, to extract and recycle the SAW energy back into the circuit.  A second IDT can be placed along the path of SAW propagation past the p-n leads.  The SAW arrives ``empty'' to the second IDT, without photogenerated electrons and holes already extracted by the p-n leads.  The traveling SAW induces an ac current in the second IDT, so acoustic energy of the SAW is transformed into ac electric energy and fed back into the first IDT to generate the SAW.  In this way, the rf energy circulates in a closed loop, and the rf generator only needs to cover incidental losses.  The electric-acoustic-electric energy conversion is commonly used in commercial SAW devices, so this technology can be adapted for a PV design.  Extinction of the SAW by the second IDT would also prevent reflection of the SAW from the terminating edge of the sample.  A standing wave, produced by two counter-propagating SAWs, is bad for electron-hole segregation, as shown in Ref.\ \cite{PRL-1997}.  Usually, reflection of the SAW is prevented by placing a damping material at the edge of the sample, but it would be more efficient to extract the SAW energy and feed it back into the loop.

Generally, active systems tend to be more efficient than passive systems, because they can be optimally and dynamically controlled.  For example, efficiency of a kitchen refrigerator is higher with an internal fan, because convection is better than diffusion, even though the fan consumes some energy.  The SAW plays a role similar to the internal fan in the proposed PV design.

\vspace{-3mm}
\paragraph*{Conclusions.}

The overall energy efficiency of the proposed device is expected to be substantially higher than for conventional PV devices.  This is because the SAW segregates electrons and holes
in a large effective volume and transports them collectively, via convection rather than diffusion.  The SAW acts as a concentrator, sweeping photogenerated electrons and holes over a wide area directly exposed to light and delivering them to a relatively small contact area.  Active management of the SAW increases efficiency of the PV process.

Wide variety of materials and techniques can be used to implement the proposed new principle.  Experiments \cite{PRL-1997,APL-2004,NatMat-2005,APL-2008,JAP-2009} were performed on GaAs with a thin layer of ZnO to enhance piezoelectric coefficient.  
Thus, it makes sense to fabricate a SAW PV prototype on GaAs to demonstrate the proof of principle.  However, the SAW idea can be also applied to cheaper PV materials, such as CdTe, CIGS, and crystalline organic semiconductors.  Potential for increasing PV efficiency in these materials using piezo- and ferro-electric effects has been pointed out \cite{Mitra-2007,Yuan-2011}, but not systematically explored and not for the SAWs.  Local electric field of the SAW can be utilized to break excitons in crystalline organic semiconductors and separate photoexcited electrons and holes, thus solving the major problem of organic PV.  Because electric charge transfer and low crystal symmetry are common for organic molecular crystals, it should be possible to identify and synthesize piezoelectric organic semiconductors.  The proposed method would not be useful for Si, because it is not piezoelectric and has an indirect energy gap.

Initial applications of the proposed SAW PV design are likely to be for powering small portable electronic devices, e.g.,\ civilian or military autonomous sensors that intermittently report their observations by radio.  These small remote devices need indefinite energy supply, but can accumulate solar energy during daytime and store it in a rechargeable  battery.  Other examples are solar-rechargeable cell phones and small consumer electronics.  Many of these devices already have rf circuitry and even use SAWs, so adding the SAW-assisted solar cells can be relatively easy.  Because these devices have small size, the highest efficiency of energy output per unit area is crucial, but price is not so crucial.  Thus, the proposed PV design would be most useful for such devices.  

The author thanks Dennis Drew and Matthew Grayson for their contributions to an earlier version arXiv:0912.5390v1 of this proposal and Ian Appelbaum for useful discussions.



\end{document}